# Equation of state and manifestation of non-spherical contribution of interaction potential in liquid rubidium metal

M.H. Ghatee[*] and F. Niroomand Hosseini

Department of Chemistry, Shiraz University, Shiraz 71454, Iran

E-mail: ghatee@susc.ac.ir (M.H. Ghatee)

Fax: +98 711 228 6008

---

[*] Corresponding author

# Equation of state and manifestation of non-spherical contribution of interaction potential in liquid rubidium metal

Abstract

A semi-empirical equation of state is presented for the liquid rubidium metal. The Lennard-Jones (8.5-4) potential model, which originally has been derived for liquid cesium metal, is found to be suitable for modeling of liquid rubidium metal as well. By applying the experimental $PVT$ data of compressed liquid metal in the range 500 K to 1600 K, the slope $B$, and intercept $C$, of the linear isotherms are determined and accordingly parameters of the potential function are calculated. The contribution of non-spherical part of the interaction potential to the second virial coefficient $B_2^{ns}$, is calculated by using the Boltzman factor that involves the proposed model potential. The multipole moments, standing as approximation of the non-spherical contribution by multipole expansion, are calculated by the Gaussian 98W program at the B3LYP level of theory. It can be concluded that the slope of the isotherm conforms to $B_2^{ns}$ quite well, though some deviations at low $T$'s exist.

# 1. Introduction

The interatomic interaction in liquid alkali metals strongly depends on the thermodynamic state of the liquid system [1]. The structure of the solid metal is usually considered to be a collection of ions, fixed in a solid matrix, creating an ionic lattice, with valence electrons delocalized over the whole lattice and, hence, electrons keep little correlation with their respective ions [2]. The ion-ion, electron-ion, and electron-electron interactions in the system are the major interactions. For liquid densities near the melting point of an alkali metal, the interatomic pair potential resemble that of solid state and may be described by pseudopotential perturbation theory based on nearly free electron model [1]. On the other hand, the metal vapor is composed of ions in addition to neutral atoms, mainly forming clusters of different sizes [1]. The forces between an atom and these clusters are of the van der Waals type forces, and may be described by the Lennard-Jones (LJ) pair potential function [3]. In fact, the interatomic interaction for alkali atoms changes from a screened columbic potential to the LJ-type interaction as the density is decreased. This leads to a rather complex interatomic interaction and therefore any theoretical method for formulation of thermophysical properties of these metals would be successful provided that an exact estimate of interatomic interaction is available.

In a series of investigations, the linear isotherm regularity (LIR) [4], which was originally devised for normal fluids has been applied to describe the thermodynamic properties of liquid alkali metals [5,6]. LIR is based on the cell theory and considers only the nearest adjacent interaction. This equation of state for a LJ (12-6) fluid states that $(Z-1)V^2$ versus $\rho^2$ is linear, where $Z = P/\rho RT$ is the compressibility factor, $\rho = 1/V$ is the molar density, $V$ is the molar volume, and $RT$ has its usual meaning. It applies highly accurate to a dense fluid and is valid for atomic, polar, nonpolar molecular systems, liquid mixtures, and quantum fluids [4]. A simple molecular model was shown to mimic the regularity to predict the temperature dependence of the intercept and the slope of $(Z-1)V^2$ versus $\rho^2$. These parameters are explicitly related to the intermolecular repulsion and attraction forces. This is because the nature of forces of LJ fluids is described well according to a definite dispersive interaction mechanism. Therefore it is of interest to investigate the detailed behavior of alkali metals in terms of a linear regularity isotherm specific to the system.

In a recent study, we have demonstrated that the slope of a different linear regularity based on the LJ (8.5-4) potential function conforms and contributes to the second virial coefficient $B_2$, of cesium metal [7]. However, its value was only a fraction of $B_2$, and on this basis we have proposed that the slope is proportional to the contribution of the non-spherical part of the potential function to the $B_2$. The validity of this proposal has been verified almost over the whole liquid range. As the usual practice, the agreement at low $T$'s is small, but at high $T$'s the agreement becomes good, and a perfect agreement is obtained, as the critical temperature is approached [7].

We will introduce a suitable potential function over the wide range of temperature to drive the linear isotherms. We apply only $PVT$ data to characterize the molecular potential parameters for liquid rubidium metal. The fact that the interionic interaction has an effect of softening on the repulsive range of the pair potential function forms the general basics for introducing the pair potential function. In this sense, rubidium is known to be harder than cesium [8]. Then we will calculate the contribution of the non-spherical interaction to the second virial coefficient of rubidium metal and demonstrate that it conforms to the slope of the corresponding linear isotherm. Multipole moments occur in the calculations, and their values are determined by *ab initio* methods.

**2.1 The isotherms**

In the previous investigations [9,10], we have shown that by modeling a system by LJ (*m-n*) potential function, one can obtain

$$(Z-1)V^{m/3} = B(1/\rho)^{(m-n)/3} + C \tag{1}$$

where $m$ and $n$ are exponents of the corresponding repulsion and attraction terms of the potential function, respectively; $B$ and $C$ are constants characteristics of the fluid system. These constants are functions of temperature and the molecular potential parameters. The forms of $B$ and $C$ are originated by the forms of attraction and repulsion of the potential function, respectively. The molecular potential parameters are obtained in terms of $B$ and $C$:

$$r_{\min} = K_{\text{bcc}}\left(\frac{-C}{B}\right)^{1/(m-n)} \tag{2}$$

and

$$\varepsilon = \left(\frac{RT}{\beta N}\right)\left(\frac{(-B)^m}{C^n}\right)^{1/(m-n)} \tag{3}$$

where $R$ is the gas constant, $N$ is the Avogadro's number, $\varepsilon$ is the potential well-depth, $r_{\min}$ is the position of the minimum potential, $K_{\text{bcc}}$ is a constant characteristics of the unit cell of the particular system, and the subscript bcc stands for body center cubic structure (assumed for liquid rubidium). For a system with body center cubic structure $K_{\mathrm{bcc}} = \sqrt{3}/(4N)^{1/3}$. $\beta$ is a constant value depending on $m$ and $n$.

The values of $B$ and $C$ can be determined from plot of the isotherm of the Eq. (1) obtained by using the corresponding experimental $PVT$ data of the liquid rubidium metal at high pressures. The accuracy of the above method for a number of normal, quantum, and metallic liquid has been shown to be quite well. Since $B$ and $C$ are temperature dependent so do $\varepsilon$ and $r_{\min}$, it turns out that the model potential function becomes temperature dependent. Therefore the method is not only accurate in the thermodynamic sense but also it provides means for the accurate determination of the effective interaction potential function.

For most practical applications, $B$ and $C$ are linear in $(1/T)$ satisfactorily. This leads to a complete determination of an equation of state for liquid alkali metal, in particular for liquid rubidium. The role of $B$ and $C$ has not been investigated fully but it has been determined that, in the case of normal liquids, the variation of $B$ conforms to $B_2$. The method of investigation and the role of $B$ have been reported recently for liquid cesium and a brief introduction will be given in the following section.

## 2.2 The non-spherical $B_2$

The slope of the $(m-n)$ isotherm $B$, is in relation with the attraction part of LJ $(m-n)$ potential function. This might accounts for the fact that $B$ conforms to $B_2$ [10]. Since the electronic structure of the metal atom is manifested by the thermodynamic properties of the liquid system $B$, therefore, the calculated $B$ may be regarded as an estimate of the interatomic electronic interaction. In this liquid metal system, the total interaction potential energy is contributed by ion-ion, ion-electron, and electron-electron interactions, as well as interparticle interaction due to dispersion forces. Generally, the dispersion of electron density of an atom could lead to an angular contribution to the interaction potential [7]. Any non-spherical interaction is contributed by the distribution of electron density resulted by the superposition of atomic orbitals. The distribution of charge density can be well-approximated by a multipole expansion of the charge density [11].

The total interaction potential energy $U$, of an $N$-atom liquid system composed of radial and angular contributions, can be represented by the sum of pair interactions using the pair potential function $u(r_{ij})$, where $r_{ij}$ is the distance between the $i$th and $j$th atoms, and the potentials of charge distribution oriented in space by the superposition of occupied atomic orbital $v(r_{ij}, \omega_i, \omega_j)$, where $\omega_i$ and $\omega_j$ are angular variables determining the orientation of $i$th and $j$th atoms:

$$U = \sum_{i>j=1}^{N} u(r_{ij}) + v(r_{ij}, \omega_i, \omega_j) \tag{4}$$

Quite obviously, the second virial coefficient of this system can be represented by the sum of radial and angular (non-spherical) contributions. Since the radial contribution is substantially lager than the angular contribution, therefore it can be treated by the perturbation method:

$$[B_2]_{ij} = [B_2^{radial}]_{ij} + [B_2^{ns}]_{ij} \tag{5}$$

where superscript *ns* and *radial* stand for non-spherical and radial contributions, respectively.

Using the (radial) pair potential function $u(r_{ij})$, the second virial coefficient due to radial contribution at a given temperature can be evaluated by numerical integration:

$$B_2^{radial} = 2\pi N \int_0^{\infty} [1 - \exp(-u(r_{ij})/k_B T)] r^2 dr \tag{6}$$

To calculate the non-spherical part of $B_2$, we follow the formulation given by Kielich [11]. (The details are shown in Appendix A.) The results show that the $B_2^{ns}$ can be calculated in terms of values of the moments appeared in the expansion, and thus we consider all effective interactions between available moments, approximated by the non-spherical charge distribution of atoms in the liquid system.

Three types of interactions contributing to $B_2^{ns}$ can be identified for rubidium based on multipole expansion (see next section and Table 1):

$$[B_2^{ns}]_{ij} = [B_2^{(\Theta-\Theta)}]_{ij} + [2B_2^{(\Theta-\Phi)}]_{ij} + [B_2^{(\Phi-\Phi)}]_{ij} \tag{7}$$

where $\Theta$ and $\Phi$ represent quadrupole and hexadecapole, respectively.

The first term on the right hand side of the Eq. (7) is due to quadrupole-quadrupole interaction ($\Theta$ - $\Theta$) with $p = 2$ and $q = 2$:

$$B_2^{(\Theta-\Theta)} = \frac{7N}{10 k_B^2 T^2} \Theta^2 \Theta^2 \int_0^{\infty} \frac{1}{r_{ij}^{10}} \exp(-\frac{u(r_{ij})}{k_B T}) r_{ij}^2 dr_{ij}. \tag{8}$$

In the same way, second term represents the quadrupole-hexadecapole interaction ($\Theta$ - $\Phi$) with $p = 2$ and $q = 4$:

$$B_2^{(\Theta-\Phi)} = \frac{22N}{4k_B^2 T^2} \Phi^2 \Theta^2 \int_0^\infty \frac{1}{r_{ij}^{14}} \exp\left(-\frac{u(r_{ij})}{k_B T}\right) r_{ij}^2 dr_{ij}. \tag{9}$$

And the last term represents hexadecapole-hexadecapole interaction ($\Phi$-$\Phi$) with $p=4$ and $q=4$:

$$B_2^{(\Phi-\Phi)} = \frac{3972N}{100 k_B^2 T^2} \Phi^2 \Phi^2 \int_0^\infty \frac{1}{r_{ij}^{18}} \exp\left(-\frac{u(r_{ij})}{k_B T}\right) r_{ij}^2 dr_{ij}. \tag{10}$$

Accordingly, by substituting the above relations in the Eq. (7),

$$B_2^{ns} = -\frac{N}{4k_B^2 T^2}\left[\frac{28}{10}\Theta^2\Theta^2\langle r_{ij}^{-10}\rangle + 22\Theta^2\Phi^2\langle r_{ij}^{-14}\rangle + \frac{3972}{25}\Phi^2\Phi^2\langle r_{ij}^{-18}\rangle\right] \tag{11}$$

## 3. Result and Discussion

For a large number of atomic and molecular fluids, LJ (12-6) potential function accounts for the pairwise interaction approximation between the fluid molecules undergoing dispersive interaction as the major interaction [4]. Liquid alkali metals have been treated thermodynamically by methods of dense normal fluids [12], in which case the structure of liquid is determined essentially by the repulsive side of the potential function. Since the single valance electrons of the two colliding alkali metal atoms overlap to form a weak chemical bond, it causes the repulsion potential becomes softer than those of normal fluids. On the other hand, alkali metal atoms in liquid state are readily polarized such that the potential function at long range is contributed by more attraction than those of normal fluids [13]. In particular the potential of a light alkali metal has a narrow and deep (hard repulsion) potential well, and a heavy alkali metal has a wide and shallow (soft repulsion) potential well.

The experimental structural studies by neutron scattering of liquid cesium as a function of temperature and pressure [14], have shown the density dependence of the effective interaction potentials of expanded alkali metals on density. These studies have shown that at high densities an alkali metal atom interacts with a soft core repulsion at small interatomic distances, and with a weak attraction at intermediate interatomic range [14,8]. The repulsive side of this potential has been analyzed in terms of an inverse power law, e.g., $a/r_{ij}^m$, where $a$ is a constant. Here for cesium, $m$=7.7, 6.8, and 5.6 at 773 K, 1373 K, and 1673 K, respectively. For rubidium, the corresponding $m$=12.4, 9.6, 9.2, and 8.3 have obtained at 350 K, 900 K, 1400 K, and 1700 K, respectively [8]. The $m$'s values for rubidium are larger, indicating the smaller the size of an alkali atom the harder the electronic cloud.

These observations have shown that the repulsive side of the potential function suitable for rubidium has a power law with $m > 6$.

In 1991, Kozhevonikov et al. [15] have used their experimental $PVT$ data of liquid cesium in the range 400 K-2000 K and have reported the internal pressure $P_{int}$. Then using these data augmented with $P_{int}$ at the proximity of absolute zero of supercooled state [16], they have derived an effective pair potential function. This potential function includes the boundary condition $P_{int} = 0$ at $r_{ij} = \sigma$ through $r_{ij} = r_{min}$ and beyond, where σ is the hard sphere diameter. From analysis of the experimental data on $P_{int}$ of the liquid cesium, the value of $m(=8.5)$ and $n(=4)$ has been determined, and thus the pair potential function $u(r_{ij}) = A\varepsilon\left[\left(\sigma/r_{ij}\right)^{8.5} - \left(\sigma/r_{ij}\right)^{4}\right]$ have been proposed to account for the interatomic interaction to model and to predict the thermodynamic properties of cesium fluid significantly accurate. [A=3.6914, $\sigma = (2.125)^{-1/4.5} r_{min}$]

By the application LJ (8.5-4) potential function (above) and employment of the previous method (section 2.1), it is predicted that the isotherm $(Z-1)V^{8.5/3}$ for cesium is an accurate linear function of $(1/\rho)^{4.5/3}$ [17], from freezing point up to the critical point. We have noticed that the same potential function is applicable to liquid rubidium, however, with much higher accuracy. In Figure 1(a), the isotherms are plotted for liquid rubidium in the range of 500 K-1600 K. To construct the isotherms, $PVT$ data has been taken from references 18 and 19. For comparison, the same isotherms calculated using LJ potential function with selected powers $m$ and $n$ are also considered (see eq. 1). As a result, the (8.5-4) isotherm presented in this work applies quite well to liquid rubidium over the whole liquid range as indicated by the linear correlation coefficient squared $(R^2)$ of the corresponding isotherms. Parameters $B$ and $C$ are rather linear functions of $(1/T)$ [see Figures 1(b) and 1(c)]. The molecular parameters $r_{min}$ and $\varepsilon$ can be calculated using the numerical values of $B$ and $C$ [see Eqs. (2) and (3)]. The $r_{min}$ increases almost linearly with temperature, a plot of which is shown in Figure 2 in the range 500 K-1600 K. By using the (8.5-4) isotherm, the values of $\varepsilon/k_B$ are calculated; these values smoothly decrease as temperature is increased [see Figure 3(a)]. The value of $\varepsilon$ in our model is actually the interaction energy of a rubidium atom with all its nearest neighboring atoms. Indeed, we have to include the values of coordination number to have a meaningful potential well-depth. To achieve this we have used the experimental coordination number reported in references 20, 21, and 22. Then the coordination number at a particular $T$ could be obtained by a smooth interpolation.

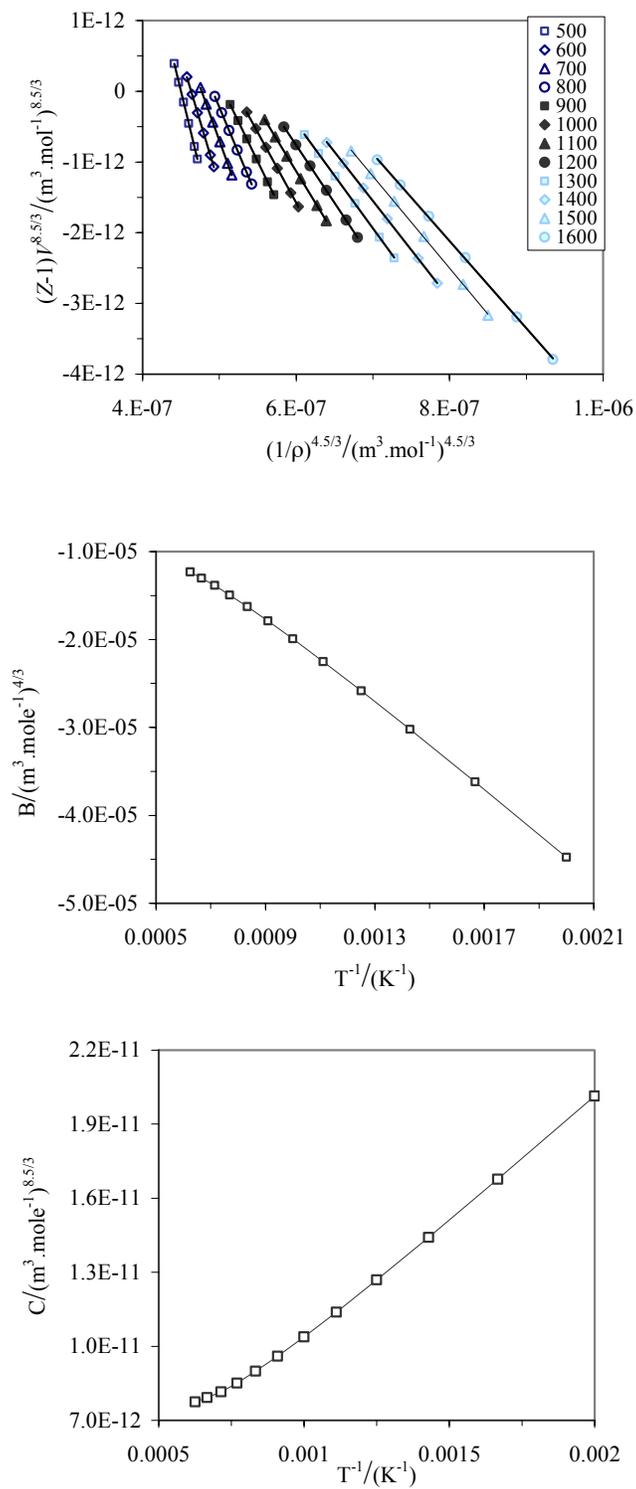

Figure 1. (a) Plots of isotherms (lines are the linear fit), (b) the values of *B*, and (c) the values of *C* (lines are trend lines).

Now, by these coordination numbers, it is seen that $\varepsilon/k_B$ smoothly decreases with temperature and almost levels off at 1100 K and increases afterward [see Figure 3(b)]. These observations are in general the same as the result for cesium except for the fact that cesium shows some wiggling at the thermodynamic states at which (second order) transitions are occurred.

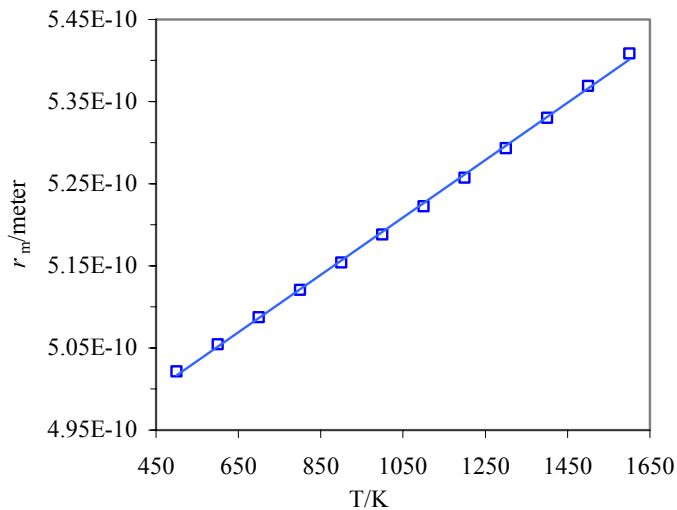

Figure 2. Plot of $r_{min}$ versus temperature for liquid rubidium metal. The line is the trend line.

So far we have demonstrated the employment of our theoretical model to derive and characterize the isotherm of the Eq.(1) for rubidium. Following the application of the characteristic potential function and consideration of the resulted parameters of the isotherm has led us to propose $B$ is equivalent to the contribution of non-spherical part of the potential function to the second virial coefficient. The (8.5-4) potential function is not angle dependent, and thus special technique must be used to calculate $B_2^{ns}$ resulting from interaction of the interatomic electron densities localized non-spherically in space as the superposition of the individual electron orbital wave functions.

To calculate the value of $B_2^{ns}$, we perform the tensor calculations, which involve the moments and the corresponding orders of the multipoles (see the Appendix A). For rubidium, $2^p - \text{pole moments}$ are determined by reference to the Gaussian 98W program [23], by applying density functional theory (B3LYP method) and different basis sets. As a result of these calculations, for quadrupole $p=2$ and for hexadecapole $p=4$. The multipoles and polarizability of the rubidium atom and the values of all components determined by different basis sets are shown in Table 1, where only the quadrupole and hexadecapole have nonzero components.

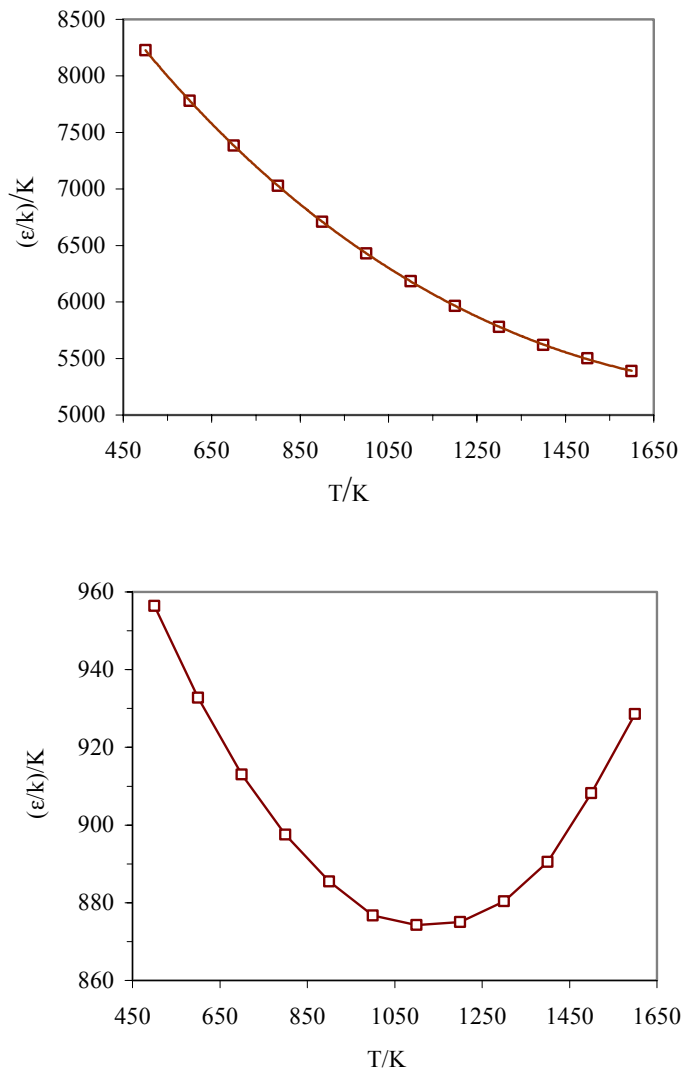

Figure 3. Plots of (a) $\varepsilon/k_B$ versus $T$ and (b) $\varepsilon/k_B$ normalized with respect to coordination number versus $T$. Lines are trend lines.

We use the output of 3-21G$^{**}$ basis set to calculate $B_2^{ns}$. The reason for selecting this basis set is its ability to predict the polarizability more accurate than the other basis sets in comparison with the available experimental polarizability. For rubidium, the experimental polarizability $\bar{\alpha}=329$ (in atomic unit) [24]. Our effort for defining (to Gaussian 98W program) specific basis set based on available exponents and coefficients data of the primitive Gaussian functions for rubidium [25], did not lead to a better polarizability.

The value $B_2^{ns}$ is determined by a two-parameter pair potential function $u(r) = f(\sigma, \varepsilon)$, which, in our method, is used to determine the isotherms defined by Eq. (1) with $m(=8.5)$ and $n(=4)$. Therefore, from Eqs. (2) and (3) it can be seen the $B_2^{ns}$ is implicitly proportional to $B$ and $C$.

**Table 1**. Components of multipoles and polarizability determined at the different levels of theory for rubidium.

| Moments[a] | Basis sets | | | | | |
|---|---|---|---|---|---|---|
| | SDD | LanL2Dz | LanL2MB | STO-3G | 3-21G** | CEP-4G |
| $q$ | 0 | 0 | 0 | 0 | 0 | 0 |
| $\mu_x, \mu_y, \mu_z$ | 0 | 0 | 0 | 0 | 0 | 0 |
| $\Theta_{xx}, \Theta_{yy}, \Theta_{zz}$ | -25.216 | -27.035 | -27.603 | -19.815 | -28.978 | -15.575 |
| $\Theta_{xy}, \Theta_{yz}, \Theta_{xz}$ | 0 | 0 | 0 | 0 | 0 | 0 |
| $\Omega_{xxx}, \Omega_{yyy}, \Omega_{zzz}, \Omega_{xxy}, \Omega_{xxz}, \Omega_{yyx}, \Omega_{yyz}, \Omega_{xzz}, \Omega_{xyz}$ | 0 | 0 | 0 | 0 | 0 | 0 |
| $\Phi_{xxxx}, \Phi_{yyyy}, \Phi_{zzzz}$ | -123.13 | -154.39 | -167.57 | -24.16 | -141.91 | -144.91 |
| $\Phi_{xxxz}, \Phi_{xxxy}, \Phi_{yyyx}, \Phi_{yyyz}, \Phi_{zzzx}, \Phi_{zzzy}$ | 0 | 0 | 0 | 0 | 0 | 0 |
| $\Phi_{xxyy}, \Phi_{yyzz}, \Phi_{yyzz}$ | -41.04 | -51.46 | -55.85 | -8.05 | -47.30 | -48.30 |
| $\Phi_{xxyz}, \Phi_{yyxz}, \Phi_{zzxy}$ | 0 | 0 | 0 | 0 | 0 | 0 |
| polarizability $\bar{\alpha} = (\alpha_{xx} + \alpha_{yy} + \alpha_{zz})/3$ | 297.27 | 345.16 | 364.60 | 20.50 | 338.57 | 400.81 |
| [a] legend of moments: $q$, Coulomb; $\mu$, Debye; $\Theta$, Debye-Ang; $\Omega$, Debye-Ang$^2$; $\Phi$, Debye-Ang$^3$; $\bar{\alpha}$, atomic unit | | | | | | |

It is important to compare $B$ with $B_2$ values calculated by integration using the LJ (8.5-4) potential function, and with $B_2^{ns}$ values obtained by the procedure of the section 2.2. These comparisons are shown in Figure 4. Notice that the value of $B$ varies smoothly and is less than experimental $B_2$ at all temperature. It can be seen that the trend of $B$ mimics the corresponding theoretical $B_2^{ns}$ value, although its absolute value is smaller at low temperatures and it comes into close agreement with $B_2^{ns}$ at high temperatures. On the other hand it is disagree with $B_2$ and its values differ by 2 to 4 orders of magnitude at all temperatures.

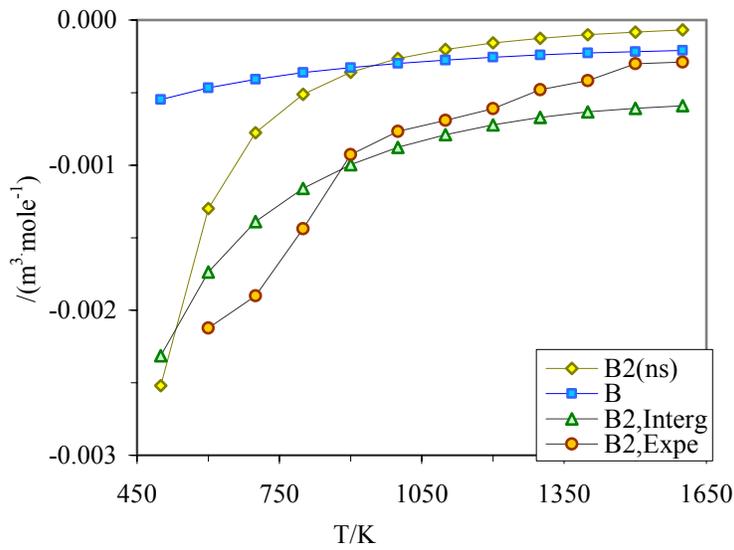

Figure 4. Comparison of $B$ with $B_2^{ns}$ as function of temperature. Experimental and theoretical $B_2$'s are shown.

The general form of the Eq. (1) allows inspection of the accuracy of a particular LJ (*m-n*) potential function for a given fluid, liquid rubidium metal in this case. In spite of the results obtained on liquid rubidium reported in reference 8, we found that LJ (8.5-4) and LJ (9-4) are the suitable potential functions as the following details. With these two potential functions, the linear isotherms are produced with excellent correlation coefficient. The potential well-depth $\varepsilon$, and the hard sphere diameter $\sigma$, are in quite agreement with experiment. Accordingly the values of $B$ versus $B_2^{ns}$ (which is of particular interest in the present study) are in good agreement at high $T$'s, while are in fair agreement at low $T$'s. With other potential, say LJ (12-6) potential, $B$ comes in closer agreement with $B_2^{ns}$ almost at all temperatures, however, no accurate correlation coefficient for the corresponding isotherms and molecular parameters can be obtained. We found it naive selecting LJ (12-6) potential as the characteristic potential function of rubidium once we further investigate prediction of transport properties such as viscosity and thermal conductivity of the vapor state of rubidium (the details and other applications to be published). For LJ (8.5-4) and LJ (9-4) potential, the absolute average deviation of viscosity is within 3.68%, and 5.36% of the experiment respectively, whereas it is within 10.38% for LJ (12-6). In the same way thermal conductivity is within 4.45%, and 6.16% of the experiment for LJ (8.5-4) and LJ (9-4) potential, respectively, whereas it is within 9.64% for LJ (12-6). The calculation for viscosity and thermal conductivity are performed by the calculation of the corresponding collision integrals in the temperature range 1000 K to 1600 K.

As stated earlier, a rather good agreement is evident at high $T$'s, whereas at low $T$'s a fair agreement exists (see Figure 4). For LJ (12-6), we have compared $B$ with $B_2^{ns}$ in a separate plot (not shown). In this case $B$ follows the trend of corresponding $B_2^{ns}$ rather perfectly and shows less deviation as compared with the case of LJ (8.5-4) potential. Comparisons are also made with $B_2$'s calculated by integration [Eq. (6)] using (8.5-4) potential function and with that of experimental data. Note that the experimental second virial coefficient is determined (in this work) by fitting the experimental data in a virial equation of state involving coefficients up to the sixth virial coefficient. It can be seen that the agreement at all temperature is good with some small deviations at low and high temperatures.

In short, the equation of state for liquid rubidium metal is determined by using a suitable potential function and its accuracy is evaluated by considering the accuracy of the derived linear isotherm. The parameters of the linear isotherm allow calculation of potential parameters from which non-spherical contribution to the second virial coefficient can be deduced.

## 4. Conclusion

Accurate equation of state for liquid rubidium metal has been derived by applying LJ (8.5-4) potential function. The accuracy of the equation of state has been substantiated by the accuracy of the linear isotherms over the whole range of liquid state where the experimental $PVT$ is available. From the slope and the intercept, parameters of the LJ (8.5-4) potential function have been determined. The multipole moments determined by the Gaussian program have been accurate enough for the perturbation calculations. By the perturbation method, it has been shown that the slope, which is related to the attraction part of the potential function, analytically conforms and numerically contributes to the non-spherical part of the second virial coefficient.

**Abbreviations**

*List of symbols*

| | |
|---|---|
| $a$ | constant |
| $A$ | constant of potential function |
| $B, C$ | constants of the equation of state |
| $B_2$ | second virial coefficient |
| $k_B$ | Boltzman factor |
| $K_{bcc}$ | unit cell constant |
| $M_i^{(p)}, M_j^{(q)}$ | $2^p$-pole and $2^q$-pole electric moments |
| $N$ | Avogadro's number |
| $P$ | pressure |
| $q$ | charge of point charge |
| $r$ | intermolecular distance |
| $R$ | gas constant |
| $R^2$ | linear correlation coefficient squared |
| $T$ | temperature |
| $u$ | pair potential |
| $U$ | total potential energy |
| $V$ | molar volume |
| $v$ | angular potential function |
| $Z$ | compression factor |
| $\langle\ \rangle$ | notation for average value |

*Greek symbols*

| | |
|---|---|
| $\alpha$ | components of polarizability |
| $\bar{\alpha}$ | average polarizability |
| $\beta$ | constant related to exponent n and m |
| $\varepsilon$ | potential well-depth |

| | | |
|---|---|---|
| $\rho$ | | molar liquid density |
| $\sigma$ | | hard-sphere diameter |
| $\omega$ | | angular orientation |
| µ | | dipole moment |
| Θ | | quadrupole moment |
| Ω | | octapole moment |
| *Ω* | | average angular part |
| Φ | | hexadecapole moment |

*Subscripts*

| | |
|---|---|
| bcc | body centered cubic |
| *i, j* | atoms |
| int | internal |
| min | minimum of potential well |
| *x, y, z* | cartesian coordinate |

*Superscript*

| | |
|---|---|
| *n, m* | exponents of the potential function |
| ns | non-spherical |
| *p, q* | order of multipole moment |
| radial | radial |
| Θ | quadrupole moment |
| Φ | hexadecapole moment |

## Acknowledgment

The authors are indebted to the Research Councils of Shiraz University for supporting this project.

## References

1. F. Hensel, Phil. Trans. R. Soc. London A 356 (1998) 97-117.

2. F. Hensel, H. Uchtman, Rev. Phys. Chem. 40 (1989) 61-68.

3. W.-c. Pilgrim, F. Hensel, J. Phys. Condens. Matter 5 (1993) B183-B192.

4. G. Parsafar, E.A. Mason, J. Phys. Chem. 97 (1993) 9048-9053.

5. M.H. Ghatee, M.H. Mousazadeh, A. Boushehri, Int. J. Thermophys. 19 (1998) 317-331.

6. E. Keshavarzi, G. Parsafar, J. Phys. Chem. B 104 (1999) 6584-6589.

7. M.H. Ghatee, F. Niroomand Hosseini, J. Phys. Chem. B 108 (2004) 10034-10040.

8. S. Munejiri, F. Shimojo, K. Hoshino, M. Watabe, J. Phys. Condens. Matter 9 (1997) 3303-3312.

9. M.H. Ghatee, H. Shams-Abadi, J. Phys. Chem. B 105 (2001) 702-710.

10. M.H. Ghatee, M. Bahadori, J. Phys. Chem. B 105 (2001) 11256-11263.

11. S. Kielich, Physica, 31 (1965) 444-460.

12. R. Redmer, H. Reinholz, G. Ropke, R. Winter, F. Noll, F. Hensel, J. Phys. Condens. Matter 4 (1992) 1659-1669.

13. R.C. Ling, J. Chem. Phys. 25 (1956) 609-612.

14. R. Winter, F. Hensel, J. Phys. Chem. 92 (1988) 7171-7174.

15. V.F. Kozhevanikov, S.P. Naurzakov, A.P. Senchankov, J. Moscow Phys. Soc. 1 (1991) 171-197.

16. I.N. Makarenko, A.M. Nikolaenko, S.M. Stishow, IOP Conf. Ser. No. 30, IOP, Bristol, 1997, p. 79.

17. M.H. Ghatee, M. Sanchooli, Fluid Phase Equilib. 214 (2003) 197-209.

18. N.B. Vargaftik, V.F. Kozhevnikov, P.N. Ermilov, V.A. Alekseev, in Proceedings of the Eighth Symposium on the Thermophysical Properties, Ed. J.V. Sengers, (New York: ASME) 1982, volume 2, p.174.

19. N.B. Vargaftik, V.F. Kozhevnikov, V.G. Stepanov, V.A. Alekseev, Yu.F. Ryzhkov, in Proceedings of the Seventh Symposium on the Thermophysical Properties, Ed. A Cezairliyan (New York:ASME) 1977, p. 927.

20. N.H. March, Phys. Chem. Liq. 20 (1989) 241-245.

21. N.H. March, J. Math. Chem. 4 (1990) 271-287.

22. N.H. March, Angel Rubio Phys. Rev. B, 56 (1997) 13865-13871.

23. M.J. Frisch, G.W. Trucks, H.B. Schlegel, G.E. Scuseria, M.A. Robb, J.R. Cheeseman, V.G. Zakrzewski, J.A. Montgomery Jr., R.E. Stratmann, J.C. Burant, S. Dapprich, J.M. Millam, A.D. Daniels, K.N. Kudin, M.C. Strain, O. Farkas, J. Tomasi, V. Barone, M. Cossi, R. Cammi, B. Mennucci, C. Pomelli, C. Adamo, S. Clifford, J. Ochterski, G.A. Petersson, P.Y. Ayala, Q. Cui, K. Morokuma, D.K. Malick, A.D. Rabuck, K. Raghavachari, J.B. Foresman, J. Cioslowski, J.V. Ortiz, A.G. Baboul, B.B. Stefanov, G. Liu, A. Liashenko, P. Piskorz, I. Komaromi, R. Gomperts, R.L. Martin, D.J. Fox, T. Keith, M.A. Al-Laham, C.Y. Peng, A. Nanayakkara, C. Gonzalez, M. Challacombe, P.M.W. Gill, B. Johnson, W. Chen, N.W. Wong, J.L. Andres, M. Head-Gordon, E.S. Replogle, J.A. Pople, Gaussian 98, Revision A.7, Gaussian, Inc., Pittsburgh, PA, **1998**.

24. Fioretti, A.; Lozeilla, J.; Massa, C.A.; Mazzoni, M.; Gabbanini, C. Optics Communications, 243 (2004) 203-208.

25. R. Poirier, R. Kari, I.G. Csizmadia,; Handbook of Gaussian basis sets; Elsevier: Amsterdam, 1985.

**Appendix A**

The atoms $i$ and $j$ can be assumed to have arbitrary charge distributions with $2^p$-pole $M_i^{(p)}$ and $2^q$-pole $M_j^{(q)}$ electric moments, respectively. The angular contribution to the second virial coefficient at temperature $T$ can be calculated by

$$[B_2^{ns}]_{ij} = -\frac{N}{2\Omega^2} \sum_{r=1}^{\infty} \frac{1}{p!}\left(\frac{-1}{k_B T}\right)^p \iiint \{v(r_{ij},\omega_i,\omega_j)\}^p \times \exp\left(-\frac{u(r_{ij})}{k_B T}\right) dr_{ij} d\omega_i d\omega_j \quad (A1)$$

where indexes $i$ and $j$ run over all atoms, and $\Omega = \int d\omega_i = \int d\omega_j$. After determining function for angular dependence of the orientation of multipole and performing the integration over the angular part of the integrand, the following relation is finally obtained:

$$[B_2^{ns}]_{ij} = -\frac{N}{6k_B^2 T^2} \sum_{p=0}^{\infty} \sum_{q=0}^{\infty} \frac{2^{p+q}(2p+2q)!(p!q!)^2}{(2p)!(2q)!(2p+1)!(2q+1)!} \times (M_i^{(p)}[r]M_i^{(p)})(M_j^{(q)}[r]M_j^{(q)})\langle r_{ij}^{-2(p+q+1)}\rangle \quad (A2)$$

where

$$\langle r_{ij}^{-2(p+q+1)}\rangle = \int r_{ij}^{-2(p+q+1)} \exp\left(-\frac{u(r_{ij})}{k_B T}\right) r_{ij}^2 dr_{ij} \quad (A3)$$

The aforementioned formulation considers a system made of a mixture of two different species $i$ and $j$. for the system consists of pure liquid rubidium metal, $i$ and $j$ are the same and, therefore, the indexes $i$ and $j$ will be dropped in the subsequent derivation.